\documentclass[floatfix,aps,showpacs,twocolumn,nofootinbib,superscriptaddress,preprintnumbers]
{revtex4}
\usepackage{graphicx}
\usepackage{amsmath}
\usepackage{subfigure}

\begin{document}

\preprint{DESY 07-004}

\title{New Constraints on Oscillations in the Primordial 
Spectrum of Inflationary Perturbations}
\author{Jan Hamann} 
\affiliation{Deutsches Elektronen-Synchrotron DESY, Notkestr. 85, 22607
    Hamburg, Germany}
\author{Laura Covi}
\affiliation{Deutsches Elektronen-Synchrotron DESY, Notkestr. 85, 22607
    Hamburg, Germany}
\author{Alessandro Melchiorri}
\affiliation{Dipartimento di Fisica and Sezione INFN,
Universit\`a di Roma ``La Sapienza'', Ple Aldo Moro 2, 00185, Italy}
\author{An\v{z}e Slosar}
\affiliation{Oxford Astrophysics, Denys Wilkinson Building, Keble Road, 
OX13RH, Oxford, United Kingdom}
\affiliation{Faculty of Mathematics and Physics, University of
  Ljubljana, Slovenia}

\date{\today}

\begin{abstract}
  We revisit the problem of constraining steps in the inflationary
  potential with cosmological data. We argue that a step in the
  inflationary potential produces qualitatively similar oscillations
  in the primordial power spectrum, independently of the details of
  the inflationary model. We propose a phenomenological description of
  these oscillations and constrain these features using a selection of
  cosmological data including the baryonic peak data from the
  correlation function of luminous red galaxies in the Sloan Digital
  Sky Survey. Our results show that degeneracies of the oscillation
  with standard cosmological parameters are virtually
  non-existent. The inclusion of new data severely tightens the
  constraints on the parameter space of oscillation parameters with
  respect to older work. This confirms that extensions to the simplest
  inflationary models can be successfully constrained using
  cosmological data.
    
\end{abstract}


\maketitle

%%%%%%%%%%%%%%%%%%%%%%%%%%%%%%%%%%%%%%%%%%%%%%%%%%%%%%%%%%%%%%%%%%%%%%%%%
%%%%%%%%%%%%%%%%%%%%%%%%%%%%%%%%%%%%%%%%%%%%%%%%%%%%%%%%%%%%%%%%%%%%%%%%%
\section{Introduction}
%%%%%%%%%%%%%%%%%%%%%%%%%%%%%%%%%%%%%%%%%%%%%%%%%%%%%%%%%%%%%%%%%%%%%%%%%
%%%%%%%%%%%%%%%%%%%%%%%%%%%%%%%%%%%%%%%%%%%%%%%%%%%%%%%%%%%%%%%%%%%%%%%%%

Recent data from the Wilkinson Microwave Anisotropy Probe (WMAP)
\cite{Spergel:2006hy,Hinshaw:2006ia,Page:2006hz,Jarosik:2006ib} 
observations of the anisotropies of the Cosmic Microwave Background
(CMB) are in excellent agreement with the predictions of inflationary
cosmology. In its simplest implementation, inflation is driven by the
potential energy of a single scalar field slowly rolling down the
potential towards the real vacuum. Under the assumption that the
potential is sufficiently flat and smooth, the resulting spectrum of
density perturbations is almost scale-invariant and can be described
with a power-law. In the context of this slow-roll paradigm, a number
of authors have used the WMAP data and various other complementary
data sets to derive bounds on the inflationary parameter space. These
include constraints on specific inflationary models
\cite{Alabidi:2006qa,Martin:2006rs,Battye:2006pk,Savage:2006tr}, the
Hubble dynamics during inflation \cite{Kinney:2006qm,Peiris:2006sj},
or, in a more empirical fashion, the parameters characterising the
primordial power spectrum
\cite{Viel:2006yh,Seljak:2006bg,Finelli:2006fi,Hamann:2006pf}.

In more general classes of inflationary models, however, slow roll may
be violated for a brief instant
\cite{Hodges:1989dw,Starobinsky:1992ts}. In single-field inflation
models, such an effect can be modelled by introducing a feature such
as a kink, bump or step \cite{Adams:2001vc} to the inflaton
potential. A step, in particular, can be regarded as an effective
field theory description of a phase transition in more realistic
multi-field models \cite{Lesgourgues:1999uc}, which may arise
naturally in, e.g., supergravity- \cite{Adams:1997de} or
M-theory-inspired inflation models \cite{Ashoorioon:2006wc}. 

This interruption of slow-roll will leave possibly detectable traces in
the primordial power spectrum. Specifically, wavelengths crossing the
horizon during this fast-roll phase will be affected
\cite{Leach:2000yw,Leach:2001zf}, leading to a deviation from the
usual power-law behaviour at these scales.
Such non-standard power spectra have been brought forward to explain
the peculiar glitches in the temperature anisotropies
\cite{Hunt:2004vt} as well as the observed low power at the largest
scales \cite{Contaldi:2003zv,Boyanovsky:2006pm}. 

Step-like features in the inflaton potential will lead to a burst of
oscillations in the primordial power spectrum. A particular
realisation of a step potential was confronted with the data in
Ref.~\cite{Peiris:2003ff} for fixed cosmological parameters
and more generally in Ref.~\cite{featureshaveafuture}.
 
In the present work, we extend the analysis of
\cite{featureshaveafuture} in several important aspects.  Firstly, we
generalise our method to spectra corresponding to a whole class of
step-inflation models with arbitrary (slow-roll) background inflaton
potentials. This allows us to derive constraints on parameters
characterising the feature in a more model-independent way.  Secondly, we
address the question of parameter degeneracies: could the presence of
a feature bias our estimates for the values and errors of the
cosmological parameters, such as the baryon or dark matter density, in
any way? Thirdly, we consider new data-sets: apart from CMB and matter
power spectrum data sets, we consider also the measurements of the
position of acoustic peak in the real space two-point galaxy
correlation function data (BAO) from the luminous red galaxy (LRG)
sample of the Sloan Digital Sky Survey (SDSS) \cite{eisenstein}.  This
data is, in principle, especially well suited to constraining (or
detecting) small amplitude oscillations in the power spectrum which
would show up as a peak in the correlation function. However, due to
biasing and weakly non-linear structure formation, this data is
difficult to interpret and we pay special attention to do it
carefully.

The paper is organised as follows: In Sec.~\ref{pert}, we briefly
remind the reader of the exact formalism of calculating the power spectrum
from a given inflaton potential and compare it with the slow-roll
approximation. In Sec.~\ref{step}, we discuss the dynamics of the
inflaton field rolling over a step and introduce a generalised step
model. Section~\ref{secCMBanalysis} is dedicated to our analysis methods
with an emphasis on the determination of the likelihood for the BAO
data set. We present our results in Sec.~\ref{results}, and, finally,
draw our conclusions in Sec.~\ref{conc}.

%%%%%%%%%%%%%%%%%%%%%%%%%%%%%%%%%%%%%%%%%%%%%%%%%%%%%%%%%%%%%%%%%%%%%%%%%
%%%%%%%%%%%%%%%%%%%%%%%%%%%%%%%%%%%%%%%%%%%%%%%%%%%%%%%%%%%%%%%%%%%%%%%%%
\section{Inflationary Perturbations}
\label{pert}
%%%%%%%%%%%%%%%%%%%%%%%%%%%%%%%%%%%%%%%%%%%%%%%%%%%%%%%%%%%%%%%%%%%%%%%%%
%%%%%%%%%%%%%%%%%%%%%%%%%%%%%%%%%%%%%%%%%%%%%%%%%%%%%%%%%%%%%%%%%%%%%%%%%

Let us start this section with a brief recapitulation of how to
calculate the primordial spectrum of curvature perturbations
$\mathcal{P_R}$, using the formalism of Stewart and Lyth
\cite{Stewart:1993bc}.

In the following, we will set $c = \hbar = 8\pi G = 1$.
We consider the gauge invariant Mukhanov variable $u$
\cite{Mukhanov:1988jd,Sasaki:1986hm} given in terms of the curvature
perturbation $\mathcal{R}$:
	\begin{equation}
		u \equiv - z \mathcal{R}.
	\end{equation}
Here, $z \equiv a \dot{\phi} / H$, where $a$ is the scale factor, $\phi$
the inflaton field, $H$ the Hubble parameter and the dot represents a
derivative with respect to time $t$.
The Fourier components of $u$ obey the equation
	\begin{equation}
	\label{uk}
		u_k''+\left( k^2 - \frac{z''}{z} \right) u_k = 0,
	\end{equation}
with a prime denoting a derivative with respect to conformal time $\tau$.

Finally, we can define the primordial power spectrum of curvature
perturbations $\mathcal{P_R}(k)$ via the two-point correlation
function
	\begin{equation}
		\langle \mathcal{R}_{k_1} \mathcal{R}_{k_2}^* \rangle
		= \frac{2 \pi^2}{k^3} \mathcal{P_R}(k) \;
		\delta^{(3)}(k_1-k_2).
	\end{equation}
Assuming gaussianity and adiabaticity, this quantity contains all
the necessary information for a complete statistical description of
the fluctuations. It is related to $u_k$ and $z$ via
	\begin{equation}
		\mathcal{P_R}(k) = \frac{k^3}{2 \pi^2} \, \left|
		\frac{u_k}{z} \right|^2.	
	\end{equation}

\subsection{Background Equations of Motion}
In order to find a solution to Equation \eqref{uk}, one needs to
know the behaviour of the term $z''/z$. Its evolution is determined
by the dynamics of the Hubble parameter and the unperturbed inflaton
field, governed by Friedmann's equation
	\begin{equation}
	\label{fe}
		H^2 =  \tfrac{1}{3} ( V +
		\tfrac{1}{2} \dot{\phi}^2), 
	\end{equation}
and the Klein-Gordon equation for $\phi$
	\begin{equation}
	\label{eom}
		\ddot{\phi} + 3 H \dot{\phi} + \frac{\text{d}
		V}{\text{d} \phi} = 0.
	\end{equation}
For our purposes, it is convenient to introduce another time
parameter, the number of $e$-foldings, defined by \mbox{$N \equiv \ln
a$}. In terms of $N$, Equations \eqref{uk}, \eqref{fe} and \eqref{eom}
read
	\begin{align}
	\label{Heq}
		&H_{,\scriptscriptstyle{N}} = -\tfrac{1}{2} H
		\phi_{,\scriptscriptstyle{N}}^2,\\ 
	\label{phieq}
		&\phi_{,\scriptscriptstyle{NN}} + \left(
		\frac{H_{,\scriptscriptstyle{N}}}{H} + 3 \right) 
		\phi_{,\scriptscriptstyle{N}} + 
		\frac{1}{H^2} \frac{\text{d}V}{\text{d}\phi} = 0,\\
	\label{ueq}
		&u_{k,\scriptscriptstyle{NN}} + \left(
		\frac{H_{,\scriptscriptstyle{N}}}{H} + 1 \right)
		u_{k,\scriptscriptstyle{N}} +\left[
		\frac{k^2}{e^{2 (N-N_0)} H^2} \vphantom{\left(
		\frac{H_{,\scriptscriptstyle{N}}}{H}\right)^2}
	        \left( 2 - \vphantom{\left(
		\frac{H_{,\scriptscriptstyle{N}}}{H}\right)^2}
		\right. \right. \\ \nonumber 
		&\left. \left.   4\,
		\frac{H_{,\scriptscriptstyle{N}}}{H}\frac{\phi_{,\scriptscriptstyle{NN}}}{\phi_{,\scriptscriptstyle{N}}} -  
		2 \left( \frac{H_{,\scriptscriptstyle{N}}}{H}
		\right)^2 - 5 \, \frac{H_{,\scriptscriptstyle{N}}}{H} -
		\frac{1}{H^2} \frac{\text{d}^2 V}{\text{d}\phi^2}
		\right) \right] u_k = 0,
	\end{align}
with $N_0$ determining the normalisation of the scale factor. This
coupled system of differential equations can easily be solved
numerically, once a suitable set of initial conditions has been
chosen.

\subsection{Initial Conditions}
Supposing that at a time $N_{sr}$ the system has reached the
inflationary attractor solution
	\begin{equation}
	\label{attr}
		\ddot{\phi} \ll 3 H \dot{\phi},
	\end{equation}
and is rolling slowly,
	\begin{equation}
	\label{sloro}
		\dot{\phi}^2 \ll V (\phi),
	\end{equation}
the initial conditions for $\phi$ and $H$ will be given by
	\begin{align}
		\phi(N_{sr}) &= \phi_{sr}\\
		 \phi_{,N}(N_{sr}) &= -\frac{1}{V(\phi_{sr})}
		 \left. \frac{\text{d} V}{\text{d}\phi}
		 \right|_{\phi_{sr}}\\
		 \quad H(\phi_{sr}) &= \sqrt{\frac{V(\phi_{sr})}{3}}.
	\end{align}
The initial conditions for $u_k$ can be obtained by requiring the late
time solution of \eqref{uk} to match the solution of a field in the
Bunch-Davies vacuum of de Sitter space, given by
	\begin{equation}
		u_k(\tau) = \frac{e^{-i k \tau}}{\sqrt{2 k}} \left(
		1 + \frac{i}{k \tau} \right),
	\end{equation}
at early times, well before the observationally relevant scales leave
the horizon. 
For $k \gg z''/z$ (or, equivalently, $k \tau \gg 1$) this can be
approximated by the free field solution in flat space
	\begin{equation}
		\label{flatsol}
		u_k = \frac{1}{\sqrt{2 k}} \, e^{-\mathrm{i}
		k \tau}.
	\end{equation}
Fixing the irrelevant phase, we obtain the initial conditions for a
mode $k$ 
	\begin{align}
		\label{ukic}
		u_k(\tau_0) &=  \frac{1}{\sqrt{2 k}},\\
		\label{dotukic}
		u_k'(\tau_0) &= -i \sqrt{\frac{k}{2}}
	\end{align}
at a time $\tau_0$ satisfying $k \gg z''/z|_{\tau_0}$.

\subsection{Slow Roll}

Following Ref. \cite{Liddle:1994dx}, we define the Hubble slow roll
parameters by
	\begin{equation}
		^n\!\beta_H \equiv \left\{ \prod^n_{i=1} \left[ -
		\frac{\text{d}\ln 
		H^{(i)}}{\text{d}\ln a} \right] \right\}^{1/n} = 2
		\left(\frac{(H^{(1)})^{n-1} H^{(n+1)}}{H^n} \right)^{1/n} 
	\end{equation}
for $n \geq 1$, with a superscript ``$(n)$'' denoting the $n\text{th}$
derivative with respect to $\phi$. In addition to that, we define
$^0\!\beta_H \equiv 2 (H^{(1)}/H)^2$. The first three parameters of the
Hubble slow roll hierarchy read
	\begin{align}
		\epsilon_H &\equiv \; ^0\beta_H = 2 \left(
		\frac{H^{(1)}(\phi)}{H(\phi)}\right)^2 = -
		\frac{\dot{H}}{H^2}, \\
		\eta_H &\equiv \; ^1\beta_H =
		\;\frac{H^{(2)}(\phi)}{H(\phi)} = -
		\frac{\ddot{\phi}}{\dot{\phi}H},\\ 
		\xi^2_H &\equiv (^2\!\beta_H)^2 = 2 \; \frac{H^{(1)}(\phi)
		H^{(3)}(\phi)}{H^2(\phi)} =
		\frac{\stackrel{...}{\phi}}{2H^2 \dot{\phi}} -
		\tfrac{1}{2} \eta_H^2.
	\end{align}

Using these definitions it can be shown that the mode equation
\eqref{uk} can be written as
	\begin{align}
		u_k''&+\left( k^2 - 2 a^2 H^2 \left[ 1+ \epsilon_H -
		\tfrac{3}{2} \eta_H + \epsilon_H^2 -
		\vphantom{\frac{1}{2}} \right. \right.\\
		\nonumber
		& \left. \left.  2 \epsilon_H
		\eta_H + \tfrac{1}{2} \eta_H^2 + \xi_H^2 \right]
		\vphantom{\frac{1}{1}} \right) u_k = 0.
	\end{align}
Note that this expression is \emph{exact}: it does not assume the
slow roll parameters to be small.

From a model-building point of view, where one regards the Lagrangian
(or the scalar potential) of the theory as the fundamental quantity,
the calculation of the Hubble slow roll parameters can be quite
involved. In this sense it may be more convenient to work with the
potential slow roll parameters instead, which use derivatives of the
potential instead of derivatives of the Hubble parameter. The first
three potential slow roll parameters are defined by
	\begin{align}
		\epsilon &\equiv \tfrac{1}{2} \left( \frac{V^{(1)}}{V}
		\right)^2,\\ 
		\eta &\equiv \frac{V^{(2)}}{V},\\
		\xi^2 &\equiv \frac{V^{(1)}V^{(3)}}{V^2}.
	\end{align}
If the attractor condition (Eq.~\eqref{attr}) is satisfied, the
two are approximately related via \cite{Liddle:1994dx}
	\begin{align}
		\epsilon_H &= \epsilon - \tfrac{4}{3} \epsilon^2 +
		\tfrac{2}{3} \epsilon \eta + \mathcal{O}(3),\\
		\eta_H &= \eta -\epsilon +\tfrac{8}{3} \epsilon^2 +
		\tfrac{1}{3} \epsilon \eta + \tfrac{1}{3} \xi^2+
		\mathcal{O}(3),\\
		\xi^2_H &= \xi^2 - 3 \epsilon \eta + 3 \epsilon^2 +
		\mathcal{O}(3),
	\end{align}
up to corrections of third and higher orders.
Expressed in terms of the potential slow roll parameters, $z''/z$ is
given by
	\begin{align}
	\label{psrzz}
		\frac{z''}{z} = & 2 a^2 H^2 \left[ 1 + \tfrac{5}{2}
		\epsilon - \tfrac{3}{2} \eta + \tfrac{7}{6} \epsilon^2
		- \right.\\ \nonumber
		&\left. \tfrac{35}{6} \epsilon \eta + \tfrac{1}{2} \eta^2 +
		\tfrac{1}{2} \xi^2 + \mathcal{O}(3) \right]
	\end{align}

It is commonly assumed that the first two slow roll parameters vary
slowly with time (i.e., $\xi_{(H)}^2 \ll 1$). Then it follows that, if
one wants to sustain inflation for long enough to solve the horizon
and flatness problems, $\epsilon_{(H)}$ and $|\eta_{(H)}|$ will also have to
be much smaller than unity. In this (``slow roll'') limit, we have
$z''/z \approx 2 a^2 H^2$, $\dot{H} \approx 0$ and $a \propto \exp [H
t]$.

Let us now turn back to Eq.~\eqref{uk}, which is basically the
equation of an oscillator with a time dependent mass term, and discuss
its solutions. The initial conditions imply that for wavenumbers with
	\begin{equation}
		\frac{k}{a} \gg H,
	\end{equation}
i.e., with wavelengths much smaller than the horizon, the solution is
given by Equation \eqref{flatsol} and $u_k$ describes a circular motion
in the complex plane. Due to the exponential growth of the scale
factor, the physical wavelenghts will be blown up and leave the
horizon, eventually satisfying \mbox{$k/a \ll H$}. In this limit, the
growing solution for $u_k$ is given by
	\begin{equation}
		u_k \propto z.
	\end{equation}

Hence, the spectrum $\mathcal{P_R}$ will converge to a constant
value for super-Hubble modes, i.e., the perturbations ``freeze
in''. We can also conclude that the fate of a perturbation with
wavelength $k$ is decided when $k/a \sim H$ and the spectrum will have
its final shape imprinted on horizon exit. It is not until much later,
when the modes reenter the horizon during radiation or matter
domination, that they will exhibit dynamical behaviour again.

Generically, the power spectrum will not be scale independent, with a
scale dependence being induced by the variation of, e.g., the
potential energy and the Hubble parameter as the inflaton field rolls down
the potential. In the slow roll regime, however, the scale dependence
is rather weak and $\mathcal{P_R}$ can be reasonably well approximated
by a power law:
	\begin{equation}
		\label{pscal}
		\mathcal{P_R}(k) \simeq A_\text{S} \left( \frac{k}{k_0}
		\right)^{n_\text{S}-1},
	\end{equation}
with the normalisation $A_\text{S}$ given by
	\begin{equation}
		A_\text{S} \simeq \left. \frac{1}{24 \pi^2}
		\frac{V}{\epsilon}\right|_{k_0=aH},
	\end{equation}
and the spectral index
	\begin{equation}
		n_\text{S} \simeq 1 - 6 \epsilon + 2 \eta.
	\end{equation}

Before we talk about relaxing some of the assumptions that went into
this analysis, let us quickly mention inflationary tensor perturbations.

\subsection{Tensor Perturbations}
Apart from the scalar perturbations described above, inflation also
generates tensor perturbations, with a spectrum given by
	\begin{equation}
		\mathcal{P}_\text{grav}(k) = \frac{k^3}{2 \pi^2} \left|
		\frac{v_k}{a} \right|^2
	\end{equation}	
and the mode equation
	\begin{equation}
		v_k'' + \left( k^2 - \frac{a''}{a} \right) v_k = 0.
	\end{equation}

This equation is very similar to the scalar one. This similarity can
be readily seen if we express the ``mass term'' $a''/a$ in terms of
the slow roll parameters:
	\begin{align}
	\label{tensorstuff}
		\frac{a''}{a} &= 2 a^2 H^2 \left[ 1 - \tfrac{1}{2}
		\epsilon_H \right]\\ 
		&\simeq 2 a^2 H^2 \left[ 1 - \tfrac{1}{2} \epsilon +
		\tfrac{2}{3} \epsilon^2 - \tfrac{1}{3} \epsilon \eta
		 + \mathcal{O}(3) \right].\nonumber
	\end{align}

Just like the scalar modes, tensor perturbations will also freeze
in at horizon exit. In the slow roll case their spectrum is
approximately
	\begin{equation}
		\label{ptens}
		\mathcal{P}_\text{grav}(k) \simeq A_\text{T} \left(
		\frac{k}{k_0} \right)^{n_\text{T}},
	\end{equation}
with the tensor spectral index
	\begin{equation}
		n_\text{T} \simeq -2 \epsilon,
	\end{equation}
and normalisation
	\begin{equation}
		A_\text{T} \simeq \left. \frac{2}{3 \pi^2} V
		\right|_{k_0 = a H}.
	\end{equation}

%%%%%%%%%%%%%%%%%%%%%%%%%%%%%%%%%%%%%%%%%%%%%%%%%%%%%%%%%%%%%%%%%%%%%%%%%
%%%%%%%%%%%%%%%%%%%%%%%%%%%%%%%%%%%%%%%%%%%%%%%%%%%%%%%%%%%%%%%%%%%%%%%%%
\section{Slow Roll Interrupted}
\label{step}
%%%%%%%%%%%%%%%%%%%%%%%%%%%%%%%%%%%%%%%%%%%%%%%%%%%%%%%%%%%%%%%%%%%%%%%%%
%%%%%%%%%%%%%%%%%%%%%%%%%%%%%%%%%%%%%%%%%%%%%%%%%%%%%%%%%%%%%%%%%%%%%%%%%

The validity of the power-law parameterisation of the primordial
spectra rests on the assumptions that the slow-roll parameters are
small and change slowly with time. Let us relax the latter and allow
$\epsilon$ and $\eta$ to change significantly on a timescale $\Delta N
\lesssim 1$. This has the consequence that we can also allow $\epsilon$
and/or $\eta$ to become of order unity momentarily, provided
that at a later time, the system returns to the slow roll regime. We
also assume here that the system starts in a state where the slow roll
conditions are fulfilled, in order to give it enough time to reach the
inflationary attractor solution.

This effect can be modelled by adding a local feature, such as a step
or a bump, to an otherwise flat inflaton potential.

\subsection{Chaotic Inflation Step Model}
Let us examine the consequences of such a feature using as an example the same
model potential as in \cite{featureshaveafuture}
	\begin{equation}
		\label{tan-potential}
		V(\phi) = \tfrac{1}{2} \, m^2 \phi^2 \,
		\left( 1 + c \tanh \left( \frac{\phi-b}{d} \right) \right).
	\end{equation}
This potential describes standard $m^2 \phi^2$ chaotic
inflation \cite{Linde:1983gd} with a step centered around $\phi =
b$. The height of the step is determined by $c$, its gradient by
$d$. We do not want inflation to be interrupted by the step, so we
stipulate $|c| \ll 1$ to ensure that the potential energy will always
dominate over the kinetic one.

\begin{figure}
	\subfigure{\includegraphics[height=.4\textwidth,angle=270]{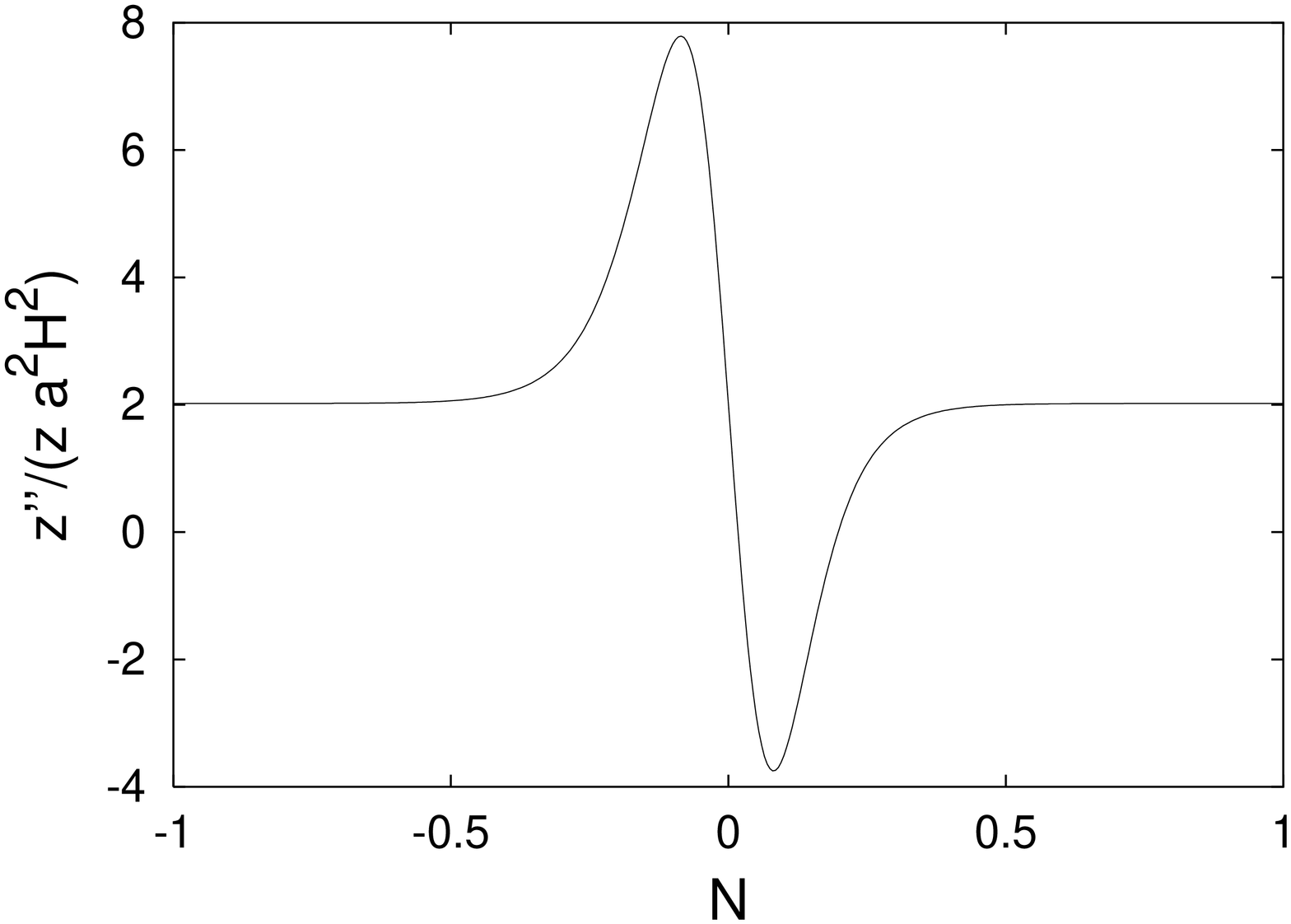}\label{zz}}  
	\subfigure{\includegraphics[height=.4\textwidth,angle=270]{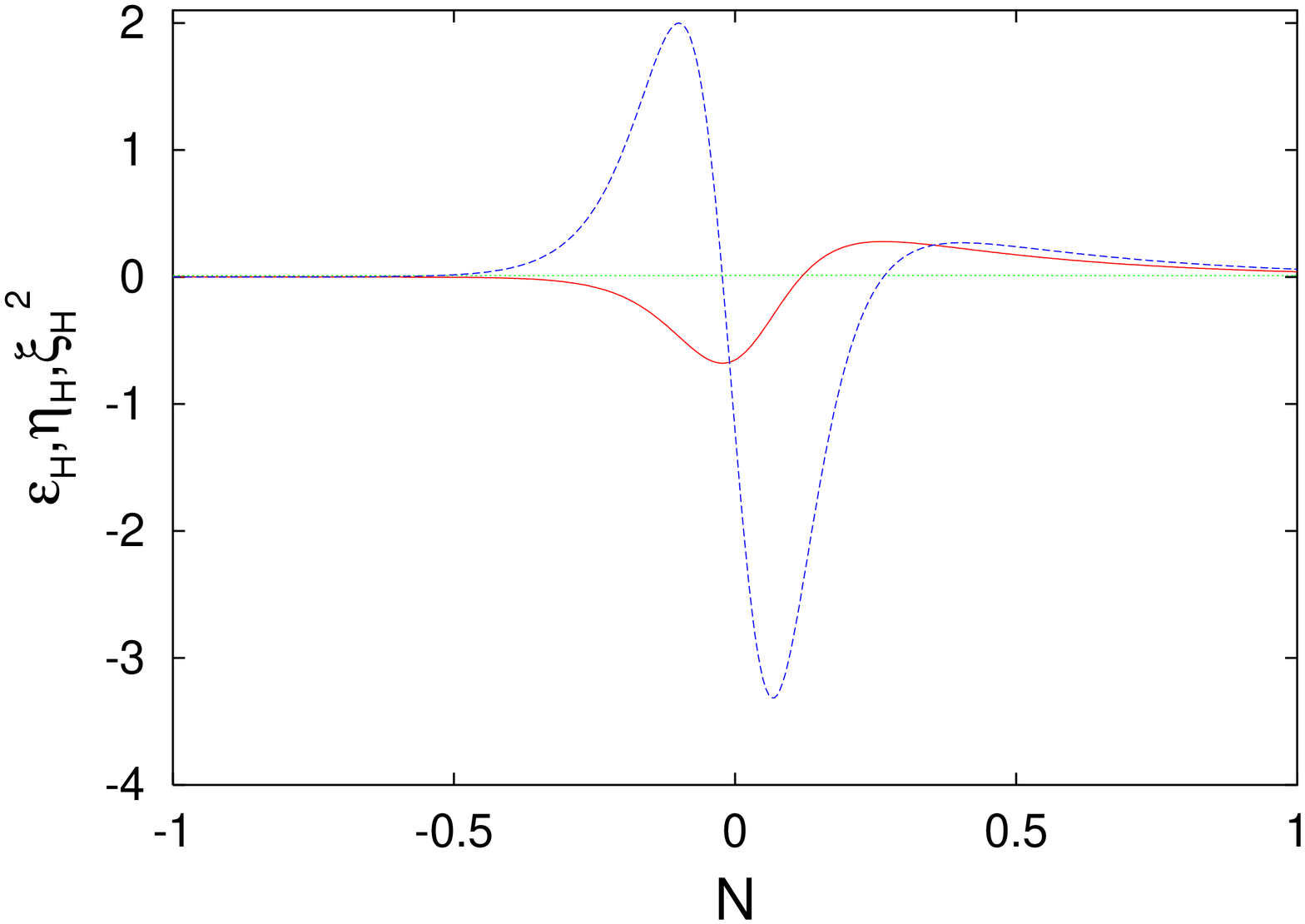}\label{hsr}}
	\caption{Top: $z''/z$ divided by $a^2 H^2$ for $b=14$,
	$c=10^{-3}$ and $d=2\times10^{-2}$ versus the number of
	$e$-foldings. $N$ is set to zero for $\phi = b$. It takes the
	inflaton field roughly half an $e$-folding to roll over the
	step.\\
	Bottom: Hubble slow roll parameters at the step,
	$\epsilon_\text{H}$ (dotted green line) remains negligible
	throughout, while $\eta_\text{H}$ (solid red line) and
	$\xi^2_\text{H}$ (dashed blue line) violate the slow roll
	conditions.}
\end{figure}

As pointed out above, the eventual spectrum crucially depends on the
dynamics of $z''/z$, which can easily be deduced from the solution of
Equations \eqref{Heq} and \eqref{phieq}. For a typical choice of
parameters, we plot the numerical solution in Figure
\ref{zz}. Generically, we find that $z''/(z a^2 H^2)$ has a maximum
before the inflaton field reaches $b$, a minimum shortly afterwards
and it will return to the asymptotic slow roll value of $\sim 2$ after
$\mathcal{O}(1)$ $e$-folding.
Comparison with the Hubble slow roll parameters (Fig.~\ref{hsr}) shows that
this behaviour is mainly caused by $\eta_\text{H}$ and
$\xi^2_\text{H}$, while $\epsilon_\text{H}$ remains small. This is a
consequence of the condition $c \ll 1$. Beware that the potential slow
roll approximation Eq.~\eqref{psrzz} will in general not work for this
potential since the contribution of higher derivative terms can be
large. The smallness of $\epsilon_H$ (and hence $\epsilon$) also
implies that there will not be any sizable deviations from a power law
for the spectrum of tensor perturbations.

\begin{figure}[h!]
\subfigure{\includegraphics[height=.23\textwidth,angle=270]{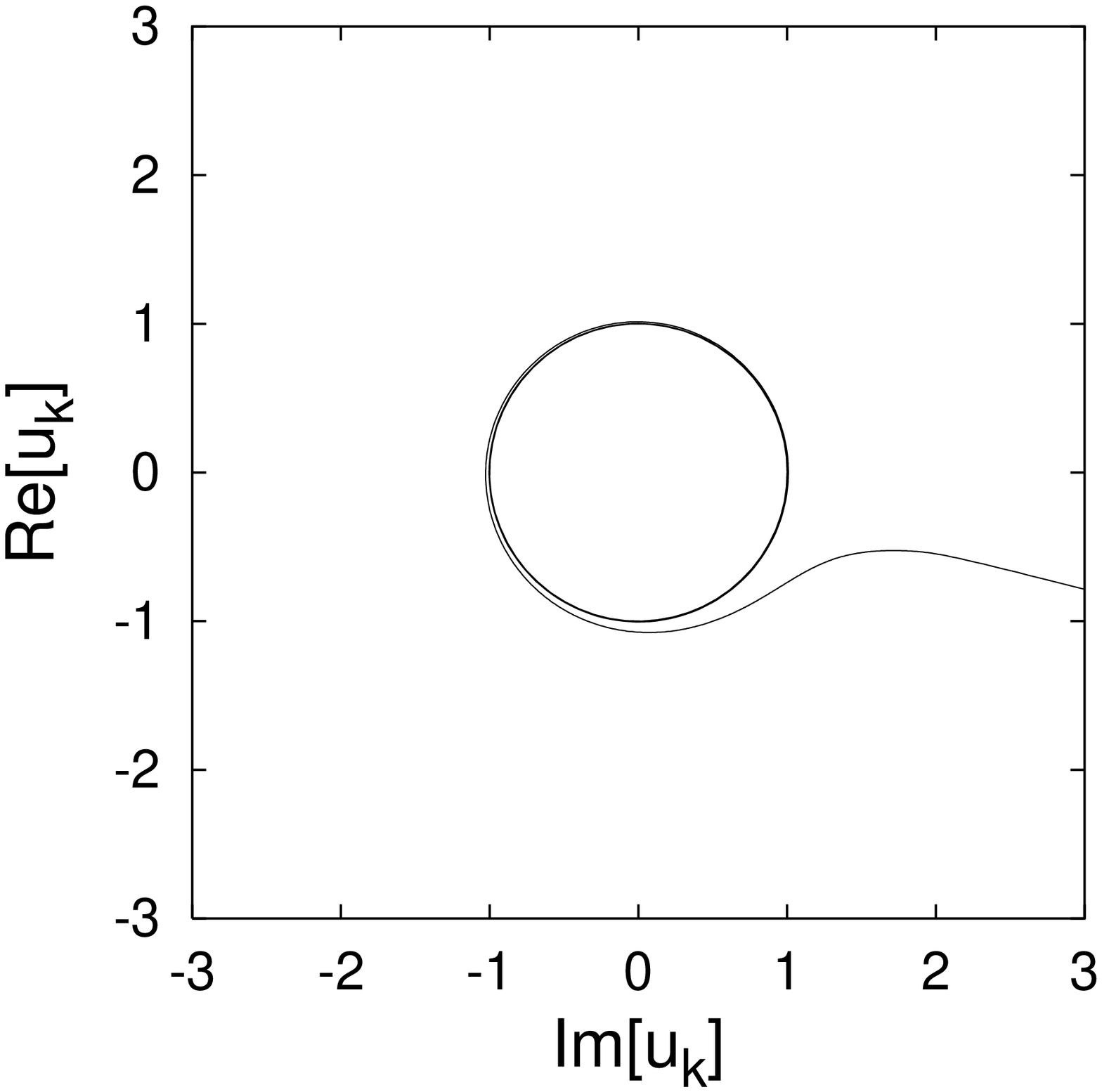}}
\subfigure{\includegraphics[height=.23\textwidth,angle=270]{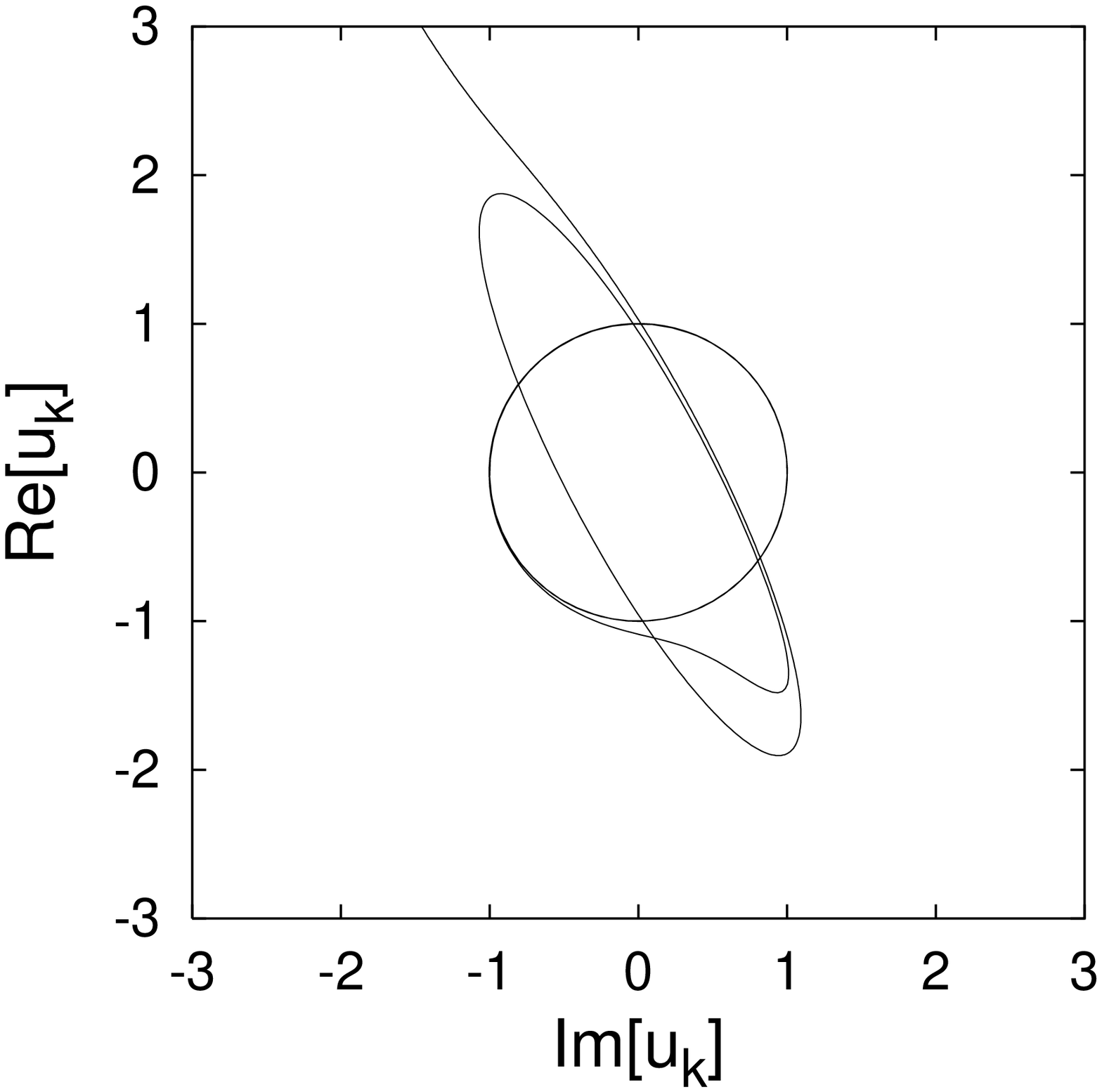}}
\subfigure{\includegraphics[height=.23\textwidth,angle=270]{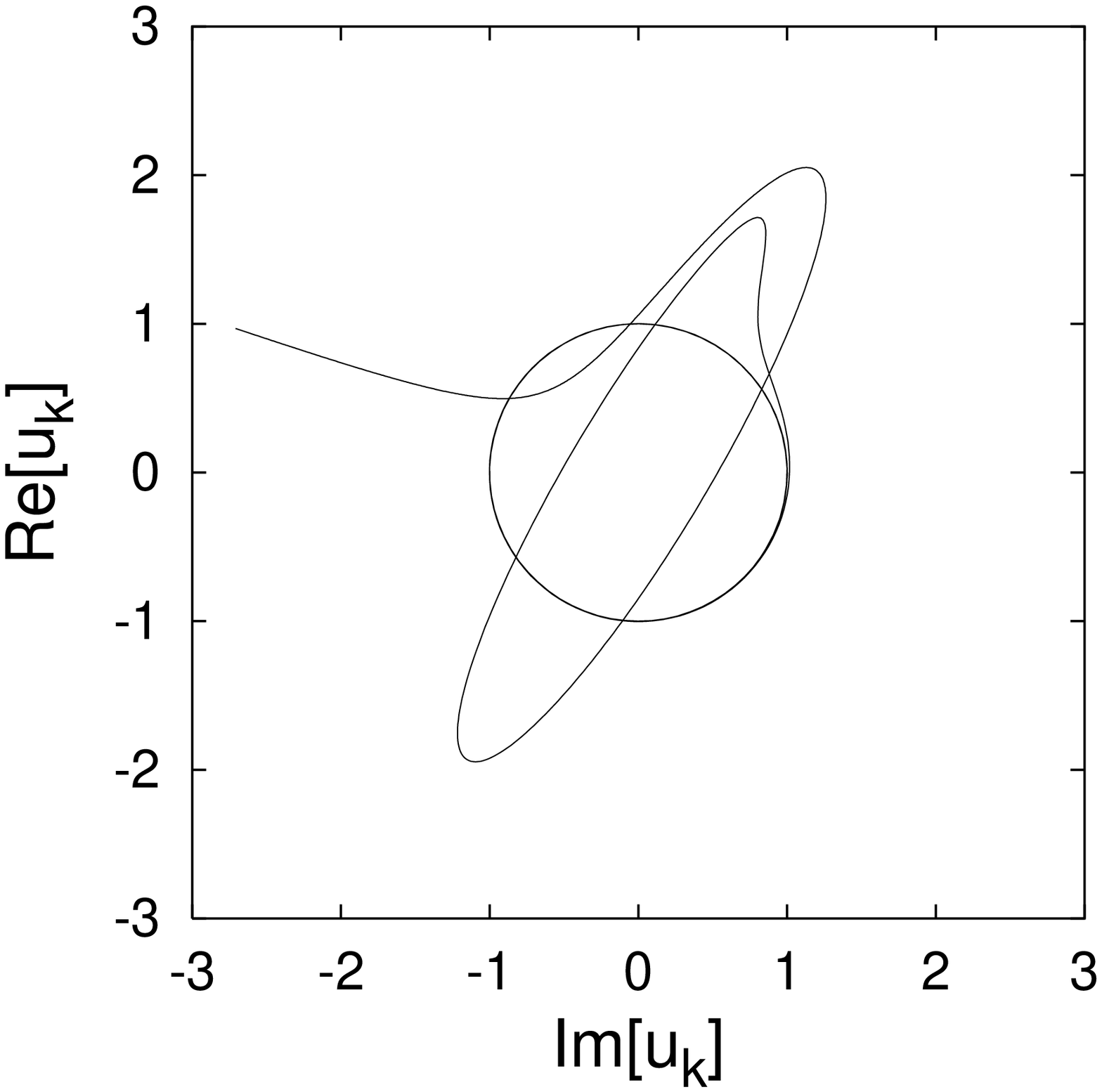}}
\caption{These figures show the evolution of $u_k$ in the complex
plane, where $u_k$ has been normalised to one in the oscillating
limit. The choice of initial conditions (\ref{ukic},\ref{dotukic})
ensures that the motion will be initially circular. The top left plot
shows a mode that is not affected by the feature, so that the circular
oscillation goes straight into a growing motion. In the other two
plots the circle gets deformed by an intermittent phase of growth
triggered by the peak of $\tfrac{z''}{z}$, to be followed by another
phase of elliptic oscillations (caused by the dip of $\tfrac{z''}{z}$)
until finally the modes leave the  horizon and start growing. Whether
a mode is suppressed or enhanced by this mechanism depends on the
phase of the oscillation when the growth sets in. Growth along the
semi-major axis will lead to an enhancement (top right), whereas
growth along the semi-minor axis entails a suppression  (bottom) with
respect to the modes of the corresponding featureless model.\label{circles}}
\end{figure}

So, how will this particular behaviour of $z''/z$ influence the
solution for $u_k$ and eventually the spectrum compared to a model
with no step? It is obvious that modes with $k^2 \gg
\text{Max}|z''/z|$, i.e., modes that are well within the horizon at
the time of the step, will not be affected at all and $u_k$ will
remain in the oscillatory regime. For $k^2 \lesssim \text{Max}|z''/z|$,
the maximum in $z''/z$ will result in a boost of exponential growth
for $u_k$, reverting to oscillations when $z''/z$ goes negative and
eventually return to the growing solution. We depict the motion of
$u_k$ in the complex plane in Fig.~\ref{circles}.

When an oscillatory phase is preceded by a growing phase, the initial
circle will be distorted to an ellipse. As the growth sets in again,
the the mode will be suppressed or enhanced, depending on the phase of
the oscillation, which itself is $k$-dependent. In the spectrum, this
can be observed as oscillations. This mechanism will be most
effective for modes that are just leaving the horizon, for modes with 
$k^2 \ll \text{Max}|z''/z|$ the phase difference will be negligible.

\begin{figure}[h!]
\includegraphics[height=.5\textwidth,angle=270]{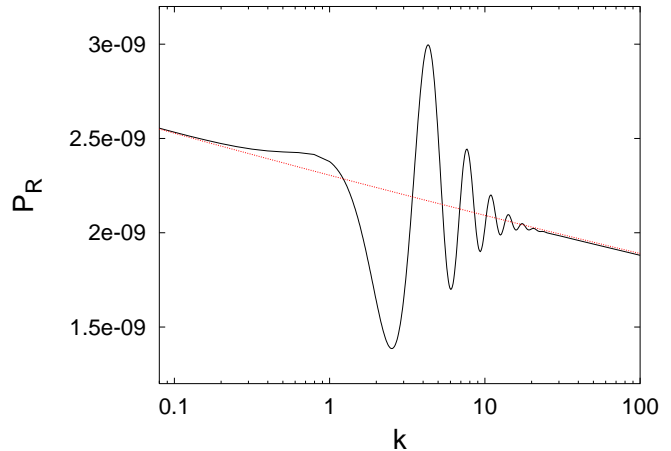}
	\caption{Primordial power spectrum for a model with \mbox{$m =
	7.5 \times 10^{-6}$}, $b=14$, $c=10^{-3}$ and
	    $d=2\times10^{-2}$ (solid black line) with wavenumber $k$
	    given in units of $a H |_{\phi=b}$. The dotted red line
	    depicts the spectrum of the same model with $c$ set to
	    zero.  \label{spectrex}}
\end{figure}

Hence, a localised feature in the potential will lead to a localised
``burst'' of oscillations in the spectrum (see also
Ref.~\cite{Burgess:2002ub}), while large and small scales will remain
unchanged with respect to the spectra of the asymptotic background
models. This is shown in Fig.~\ref{spectrex}. 
Note that the wavelengths affected by the feature are those that are
about to leave the horizon as the inflaton field reaches the centre of
the step. In particular, also the frequency of the oscillations of the
spectrum is proportional to this scale.

What remains is to identify the horizon size at the step with a
physical scale today. This connection can be made if one knows the
total number of $e$-foldings $N_*$ of inflation that took place after
a known physical scale $k_*$ left the horizon.
Technically, we evolve the background equations
(\ref{Heq},\ref{phieq}) until the end of inflation $N_\text{end}$,
(defined by $\ddot{a}(N_\text{end}) = 0$). The scale $k_*$ can then be
determined in units of $a H |_{\phi=b}$ via
	\begin{equation}
		k_* \leftrightarrow \frac{a(N_\text{end}-N_*) \;
		H(N_\text{end}-N_*)}{a H |_{\phi=b}}.
	\end{equation}
As long as the spectrum of the $c=0$ background model is only mildly
scale dependent, there will be a strong degeneracy between $N_*$ and
$b$: shifting the feature in the potential will have the same effect
as shifting the scale of $k$. In the following we will therefore not
treat $N_*$ as a free parameter, but set $N_* =50$ for $k_* = 0.05 \;
h \,\text{Mpc}^{-1}$. If we want the feature to affect scales that are
within reach of current observations, this will require $b$ to lie in
the interval $14 \lesssim b \lesssim 15$.

\subsection{Model Dependence}

Having analysed a specific example in the previous subsection, let
us now address the question of model dependence: Will we arrive at
different conclusions if we modify the background inflationary model
(e.g., $\lambda \phi^4$ instead of $m^2 \phi^2$) or the
parameterisation of the step?

We will argue that a more general potential
	\begin{equation}
	\label{vgen}
		V(\phi) = V_0 + f(\phi) \; S(\phi-b)
	\end{equation}
leads to a qualitatively similar spectrum as the potential
\eqref{tan-potential}. Here, $V_\text{bg}(\phi) \equiv
V_0+f(\phi)$ is the background potential, which should fulfil the
slow roll conditions with $f$ and $V_0$ positive definite. The
function $S(\phi)$ parameterises the step, and should monotonically
asymptote to $1\pm c$ ($c \ll 1$) for $\phi \gg b$ and $\phi \ll b$,
respectively, with $S(0) = 1$.

As we have seen above, the derivatives of the potential are crucial to
determining the spectrum. In general, the derivatives of $V$ are given
by
	\begin{equation}
		V^{(n)}(\phi) = \sum_{i=0}^n \binom{n}{i}
		f^{(i)}(\phi) \; S^{(n-i)}(\phi).
	\end{equation}

Far away from the step, the derivatives of $S$ will be negligible and
the potential and its derivatives are approximately
	\begin{align}
		&V(\phi) \simeq V_0 + f(\phi) (1 \pm c) \simeq
		V_\text{bg}(\phi),\\
		&V^{(n)}(\phi) \simeq f^{(n)}(\phi) (1 \pm c)
		\simeq V_{\text{bg}}^{(n)}(\phi).
	\end{align}
Since the slow roll conditions hold here, the spectrum will be given
by Eq. \eqref{pscal} with
	\begin{align}
		A_\text{S} &\simeq A_\text{S}^\text{bg} \left(1\pm c
		\left( \frac{3f}{V_0+f} -2 \right) + \mathcal{O}(c^2)
		\right),\\
		n_\text{S} &\simeq n_\text{S}^\text{bg} \pm c
		\left( \frac{2 V_0}{V_0+f} \left( \eta^\text{bg} - 6
		\epsilon^\text{bg} \right) \right)+ \mathcal{O}(c^2).
	\end{align}
In the special case $V_0 = 0$, we have exactly \mbox{$A_\text{S} =
A_\text{S}^\text{bg} (1\pm c)$} and \mbox{$n_\text{S} =
n_\text{S}^\text{bg}$}. If $V_0 \neq 0$, there are additional
corrections of order $c$ to the normalisation and also corrections to
the tilt, which are suppressed by $c$ and the slow-roll parameters of
the background model. In both cases, one asymptotically recovers the
spectrum of the background model in the limit $c \ll 1$. 

\begin{figure}[h!]
\includegraphics[height=.5\textwidth,angle=270]{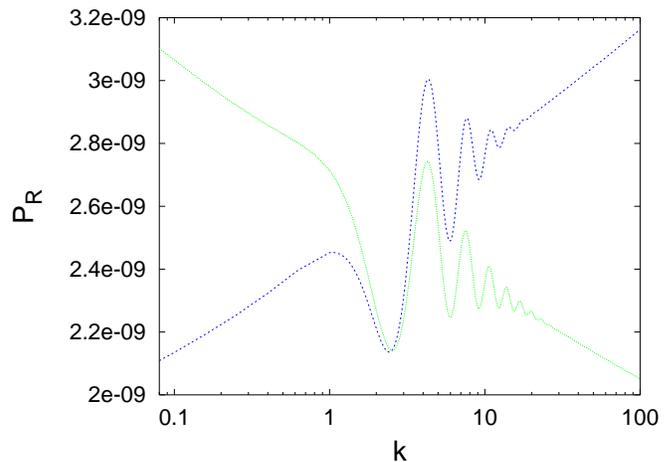}
	\caption{Primordial power spectra of a hybrid inflation type
	step model \eqref{hybridpot} with $V_0= 3.7 \times 10^{-14}$,
	$m=3.2 \times 10^{-8}$, $b = 0.0125$, $c = 10^{-3}$ and $d=5
	\times 10^{-5}$ (dashed blue line), and of potential
	\eqref{phi4pot} with parameters $\lambda = 6 \times
	10^{-14}$, $b=21$, $c=5 \times 10^{-4}$ and $d=0.02$ (dotted
	green line). The hybrid inflation background model has
	$n_\text{S} > 1$, suppressing large scale fluctuations, while
	the $\lambda \phi^4$ model has $n_\text{S} < 1$ with more
	power on large scales.
	\label{funnyspectra}}
\end{figure}

Near the step, however, the derivatives of $V$ will have a
contribution from the derivatives of $S$. If the step is sharp enough,
the $n$th derivative of $V$ will be dominated by the $n$th derivative
of $S$, since the other terms are suppressed with factors of the order
of the slow roll parameters of the background model. Hence,
the dynamics of $z''/z$ near the step hardly depends on the
background, but is determined by the form of $S$. On the other hand,
any $S$ that gives a $z''/z$ which roughly shows a behaviour like the
one depicted in Fig.~\ref{zz}, will lead to a burst of oscillations
in the power spectrum. The similarities between spectra of different
background models are illustrated in Fig.~\ref{funnyspectra}, where
we plot the spectra of a hybrid inflation type potential
	\begin{equation}
	\label{hybridpot}
		V(\phi) = V_0 + \tfrac{1}{2} \, m^2 \phi^2 \,
		\left( 1 + c \tanh \left( \frac{\phi-b}{d} \right) \right),
	\end{equation}
and another monomial potential with a different form of the step
function
	\begin{equation}
	\label{phi4pot}
		V(\phi) =  \lambda \, \phi^4 \,
		\left( 1 + c \arctan \left( \frac{\phi-b}{d} \right) \right).
	\end{equation}
Note that despite the difference in background models and step
functions, the maxima and minima of the oscillations occur at the same
wavelengths.

To alleviate the model dependence of the analysis when confronting
theory with experiment, we choose a phenomenological approach and
define the spectrum of a generalised step model
	\begin{equation}
	\label{gsmspec}
		\mathcal{P_R}^\text{gsm} = \mathcal{P_R}^\text{step}
		\left( \frac{k}{k_0}
		\right)^{n_\text{S}-n_\text{S}^\text{step}}.
	\end{equation}
Here, $\mathcal{P_R}^\text{step}$ is the spectrum obtained from the
potential Eq.~\eqref{tan-potential} and $n_\text{S}^\text{step} =
0.96$ is the spectral index of the $\tfrac{1}{2} m^2 \phi^2$
model. The quantity $n_\text{S}$ then describes the overall effective
tilt of the spectrum. Spectra of this type will arise from potentials
of the form Eq.~\eqref{vgen}. While the fine details of particular
models may differ slightly from this approximation,
Eq.~\eqref{gsmspec} will nevertheless capture the broad features of a
large class of background models, since, as argued above, the shape of
the burst of oscillations is largely independent of the background
model. Minor differences would likely be washed out in the angular
power spectrum of the CMB anyway \cite{Kawasaki:2004pi}. The
asymptotic behaviour of models with $V_0 = 0$ will be reproduced
exactly; for $V_0 > 0$ it will be approximate, with errors of order $c$.

There is a catch however: in this analysis the parameters $b$, $c$ and
$d$ will be bereaved of their meaning as parameters of the
potential. Instead, they should be interpreted as phenomenological
parameters which describe the spectrum. This does not preclude us from
deriving meaningful constraints, though. We argued that the shape of
the modulation of the spectrum is largely independent of the
background, so similar modulations should be the consequence of
similar step dynamics. A useful quantity in this context is the
maximum value the slow roll parameters $\epsilon$, $\eta$ and $\xi^2$
can reach at the step. For the potential \eqref{tan-potential}, we can
estimate $\epsilon_\text{max}$, $\eta_\text{max}$ and
$\xi^2_\text{max}$ in terms of $b$, $c$ and $d$:
	\begin{align}
	\label{eps}
		\epsilon_\text{max} &\simeq \epsilon^\text{bg} +
		\frac{c^2}{2 d^2} + \frac{2 c}{b d},\\
	\label{eta}
		\eta_\text{max} &\simeq  \eta^\text{bg} + 0.77
		\frac{c}{d^2},\\
	\label{xi}
		|\xi^2_\text{max}| &\simeq 2 \frac{c^2}{d^4} + 4
		\frac{c}{b d^3},
	\end{align}
assuming $c < 1$, $d < 1$ and $b > 1$. Note that $\xi^2 = 0$ for the
background model.

Along the same lines, one can can replace $b$ with $k_\text{s}$,
corresponding to today's wavenumber of the perturbations that left the
horizon during inflation when $\phi = b$.

%%%%%%%%%%%%%%%%%%%%%%%%%%%%%%%%%%%%%%%%%%%%%%%%%%%%%%%%%%%%%%%%%%%%%%%%%
%%%%%%%%%%%%%%%%%%%%%%%%%%%%%%%%%%%%%%%%%%%%%%%%%%%%%%%%%%%%%%%%%%%%%%%%%
\section{Data Analysis}
\label{secCMBanalysis}
%%%%%%%%%%%%%%%%%%%%%%%%%%%%%%%%%%%%%%%%%%%%%%%%%%%%%%%%%%%%%%%%%%%%%%%%%
%%%%%%%%%%%%%%%%%%%%%%%%%%%%%%%%%%%%%%%%%%%%%%%%%%%%%%%%%%%%%%%%%%%%%%%%%

We compare the theoretical predictions of three theoretical models
(A, B and C) with observational data. We use the Markov chain Monte
Carlo (MCMC) package \texttt{cosmomc} \cite{cosmomc} to reconstruct
the likelihood function in the space of model parameters and infer
constraints on these parameters.

\subsection{Models}
The three models have four parameters in common: $\omega_b$
(baryon density), $\omega_c$ (CDM density), $\tau$ (optical depth to
reionisation) and $\theta_s$ (sound horizon/angular diameter distance
at decoupling). The difference lies in the primordial power spectrum.

\begin{itemize}
	\item[A.]{Vanilla power-law $\Lambda$CDM model: the initial
	spectrum is parameterised with $A_\text{S}$ and $n_\text{S}$}.
	\item[B.]{Step model (Eq.~\eqref{tan-potential}) with
	parameters $A_\text{S}$, $b$, $c$ and $d$.}
	\item[C.]{Generalised step model, which uses an effective tilt
	$n_\text{S}$ in addition to the parameters of model
	B. Constraints on  $\epsilon_\text{max} - \epsilon^\text{bg}$,
	$\eta_\text{max} - \eta^\text{bg}$ and $\xi^2_\text{max}$ are
	derived using Eqns.~(\ref{eps})-(\ref{xi}).}
\end{itemize}
We limit our analysis to scalar perturbations. While tensor
perturbations may, in principle, give a subdominant contribution,
their spectrum will be smooth in the class of models studied here, so
we do not expect any major degeneracies with the step parameters.

\subsection{Data Sets}
To assess the influence of different data on the constraints, we
perform the analysis for each of the models using three different sets
of data:

\begin{itemize}
	\item[1.]{WMAP three year temperature and polarisation
	anisotropy
	data\cite{Spergel:2006hy,Hinshaw:2006ia,Page:2006hz,Jarosik:2006ib}
	(WMAP3). The likelihood is determined using the October 2006
	version of the WMAP likelihood code available at the \texttt{LAMBDA}
	website \cite{lambda}.}
	\item[2.]{WMAP3 plus small scale CMB temperature anisotropy
	data from the ACBAR \cite{acbar}, BOOMERANG \cite{boomerang}
	and CBI \cite{cbi} experiments, plus the power spectrum data
	of the luminous red galaxy sample from the Sloan Digital Sky
	Survey (SDSS), data release 4 \cite{Tegmark:2006az}. To avoid
	a dependence of our results on nonlinear modelling, we only
	use the first 13 $k$-bands ($k/h < 0.09$ Mpc$^{-1}$).}
	\item[3.]{Same as data set 2, plus two-point correlation
	function data from the SDSS LRG \cite{eisenstein}.}
\end{itemize}

\subsection{Analysis}
Our constraints are derived from eight parallel chains generated using
the Metropolis algorithm \cite{Metropolis:1953am}. We use the Gelman
and Rubin $R$ parameter \cite{gelru} to keep track of convergence of
the chains, stopping the chains at \mbox{$R - 1 < 0.05$}. Since the
likelihood function is highly nongaussian in some parameter directions
and even multimodal in certain cases, we double-check our results by
comparing with chains generated with a variation of the
multicanonical sampling algorithm \cite{Berg:1998nj}.

\subsection{Priors}

Apart from the hard-coded priors of \texttt{cosmomc} on $H_0$ ($
40$ km/s/Mpc $< H_0< 100$ km/s/Mpc) and the age of the Universe
($10^{10}$ a $< A < 2 \times 10^{10}$ a), we impose flat priors on the
other cosmological parameters. For the parameters of the potential we
choose a flat prior on \mbox{$b \in [14,15]$} and logarithmic priors
on $c,d$ and $c/d^2$ (\mbox{$\log c \in [-6,-1]$}, \mbox{$\log d \in
[-2.5, -0.5]$}, \mbox{$\log c/d^2 \in [-5,3]$}).

\subsection{Baryon Acoustic Peak}
\label{bao}
Oscillations in the dark matter power spectrum due to acoustic
oscillations in the plasma prior to decoupling result in an single
peak in the two-point correlation function of the distribution of
galaxies $\xi(r)$. In Ref.~\cite{eisenstein}, the authors claim the
detection of such a peak and identify it as corresponding to the
baryonic oscillations of the matter power spectrum. \\
Since any oscillation of the spectrum, regardless of its origin, will
lead to a feature in the correlation function, this data set is
particularly well suited to constraining oscillations in the initial
power spectrum as well, provided that the features are not completely
washed out through subsequent evolution.

The correlation function is related to the matter power spectrum
$P(k)$ via a Fourier transform:
\begin{equation}
\xi(r) \propto \int_0^\infty \text{d} k \,k^2 P(k) \frac{\sin kr}{kr}.
\end{equation}
Technically, the upper limit of the integral would be some ultraviolet
cutoff $k_{UV}$, chosen such that the error in $\xi$ is small ($\ll
1\%$). For the scales covered by the SDSS data, i.e., comoving
separations between 12 and 175 $h^{-1} \text{Mpc}$, this requires a
momentum cutoff $k_{UV} > 1 \, h/\text{Mpc}$. At these wavenumbers,
however, nonlinear effects cannot be neglected anymore, which makes
the theoretical prediction of $\xi$ somewhat tricky.

The standard procedure is outlined in section 4.2 of
Ref.~\cite{eisenstein} and involves corrections for redshift space
distortion, nonlinear clustering, scale dependent bias, and a
smoothing of features on small scales due to mode coupling. All of
these methods were calibrated with nonlinear simulations in a vanilla
cosmology setting and it is not obvious that they should be applicable
to our case. With the exception of the smoothing, however, the effect
of these corrections on the correlation function is smaller than 10\%
and will only be noticable at scales $< 40 h^{-1}$ Mpc (see Figure 5
of \cite{eisenstein}). So even if we assume a large uncertainty in the
nonlinear corrections, the accuracy of the theoretical correlation
function will still be of order a few per cent, that is smaller than the
error bars of the data.

Let us look at the smoothing procedure in a bit more detail. In the
usual case, the dewiggled transfer function $T_{\text{dw}}$ is a
weighted interpolation between the linear transfer function
$T_{\text{lin}}$ and the Eisenstein-Hu \cite{eihu} no-wiggle transfer
function $T_{\text{nw}}$
\begin{equation}
T_{\text{dw}}(k) = w(k) \, T_{\text{lin}}(k)  + \left( 1 - w(k)
\right) \,
T_{\text{nw}},
\end{equation}
with a weight function $w(k) = \exp \left[ - (a k)^2 \right]$ and
\mbox{$a = 7 h^{-1}$} Mpc. 
This is related to the dewiggled spectrum by
\begin{equation}
P_{\text{dw}}(k) = k \, T_{\text{dw}}^2(k) \, \mathcal{P_R}(k).
\end{equation}

In the case of a non-smooth primordial power spectrum
$\mathcal{P_R}(k)$, one should of course also dewiggle the initial
features. In order to recover the standard procedure for power-law
spectra, we will instead smooth the quantity 
\begin{equation}
\hat{T}(k) \equiv \left( P(k)/k \right)^{1/2} = T(k) \sqrt{
\mathcal{P_R}(k)}.
\end{equation}

The use of the no-wiggle transfer function rests on the assumption
that at small scales, mode coupling will totally erase all structure,
which is reasonable as long as the amplitude of features is of the
same order as that of the baryon oscillations. For much larger
oscillations, mode coupling might not be efficient enough to erase all
structure; it is likely that some residual oscillations will
remain. So instead of a no-wiggle $\hat{T}_{\text{nw}}$, we will use a
smoothed $\hat{T}_\text{s}$ defined by 
\begin{equation}
\hat{T}_{\text{s}}(k,q) = \exp \left[ \frac{1}{q} \int_{\ln k-q/2}^{\ln k+q/2}
\text{d}\!\ln\!k' \, \ln \left[ \hat{T}_{\text{lin}}(k') \right] \right],
\end{equation}
i.e., a convolution of $\hat{T}_\text{lin}$ with a top hat function of
width $q$ in log-log space. 
The dewiggled power spectrum is then given by
\begin{equation}
P_{\text{dw}}(k,q) = k \left( w(k) \, \hat{T}_{\text{lin}}(k)  +
\left( 1 - w(k) \right) \, \hat{T}_{\text{s}}(k,q) \right)^2.
\end{equation}

Without turning to $N$-body simulations it would be hard to estimate
how much the spectrum will have to be smoothed, though. Therefore, we
will determine the BAO likelihood $\mathcal{L}_\text{BAO}$ by
marginalising over $q$:
\begin{equation}
\label{baolike}
\mathcal{L}_\text{BAO} = \int \text{d}q \, \mathcal{L}(q) \pi (q).
\end{equation}
%We take the prior $\pi(q)$ to be a top hat function between $q = 0$
%(i.e., no smoothing at all) and an upper value $q_\text{max}$, chosen
%such that it lies in a region where $\mathcal{L}(q)$ is flat in $q$,
%corresponding to a complete smoothing.
%In our numerical code, we approximate the integral \eqref{baolike} by
%averaging the likelihood over $N$ values of $q$:
%\begin{equation}
%-\ln \mathcal{L}_\text{BAO} \simeq \text{Min}(- \ln \mathcal{L}(q_i)) - \ln
%\left[ \frac{1}{N} \sum_{i=1}^N \frac{\mathcal{L}(q_i)}{\text{Max}(\mathcal{L}(q_i))} \right],
%\end{equation}
%with $q_i = (i-1) \, q_\text{max}/(N-1)$.

%%%%%%%%%%%%%%%%%%%%%%%%%%%%%%%%%%%%%%%%%%%%%%%%%%%%%%%%%%%%%%%%%%%%%%%%%
%%%%%%%%%%%%%%%%%%%%%%%%%%%%%%%%%%%%%%%%%%%%%%%%%%%%%%%%%%%%%%%%%%%%%%%%%
\section{Results}
\label{results}
%%%%%%%%%%%%%%%%%%%%%%%%%%%%%%%%%%%%%%%%%%%%%%%%%%%%%%%%%%%%%%%%%%%%%%%%%
%%%%%%%%%%%%%%%%%%%%%%%%%%%%%%%%%%%%%%%%%%%%%%%%%%%%%%%%%%%%%%%%%%%%%%%%%

\begin{figure}[!ht]
\includegraphics[width=.5\textwidth]{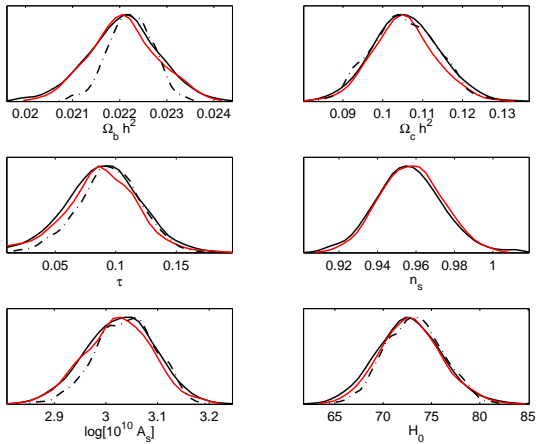}
	\caption{Marginalised likelihoods for model A (solid red
	line), model B (dot-dashed line) and model C (solid black
	line). The differences between the results for A and C are
	marginal. For some parameters, the results for model B differ
	slightly. This should be attributed to the degeneracies of the
	spectral index with these parameters and the fact that the
	tilt of the spectrum is fixed in this model, it is not due to
	the presence of a feature.
	\label{1dmarge}}
\end{figure}

An important question in the context of a model-dependent analysis is
how the choice of model will affect the estimates of the
parameters, particularly if the models are nested. Possible
degeneracies between ``standard'' and newly introduced parameters can
bias means as well as errors. In Fig.~\ref{1dmarge}, we plot the
marginalised likelihood distributions for the vanilla parameters for
all three models with data set 1. There are small differences between
models A and B for $\Omega_b h^2$, $\tau$ and the normalisation. These
arise due to the fact that in model B, the tilt of the spectrum is
fixed. There is a well-known degeneracy between these parameters and
the spectral index. Fixing the tilt near the best fit value will
reduce the errors on the parameters it is degenerate with, which is
precisely what is happening here.

The distributions for models A and C show a remarkable similarity
which leads us to conclude that the presence of a feature will not
have any statistically significant influence on the results for the
parameters of the vanilla model. This conclusion remains unchanged if
we consider the other data sets.

Another interesting question is whether the data prefer the presence
of a feature over a smooth spectrum. How much will a feature improve
the fit and can we understand why?

In Ref.~\cite{featureshaveafuture}, we studied model B and found two
regions in parameter space which improve the fit to the WMAP3 data by
$\Delta \chi^2 \sim 5$ and $\Delta \chi^2 \sim 7$, respectively. The
former corresponds to oscillations at large scales ($\ell \simeq
20-30$), while the latter has oscillations of a wavelength similar to
the baryonic acoustic oscillations and lies near the third peak of the
CMB temperature power spectrum. Adding small scale CMB data and, in
particular, the power spectrum data of the 2003 data release of
the SDSS \cite{Tegmark:2003uf}, improved the fit of the small scale
maximum to $\Delta \chi^2 \sim 15$.

\begin{figure}[!ht]
	\includegraphics[width=.4\textwidth]{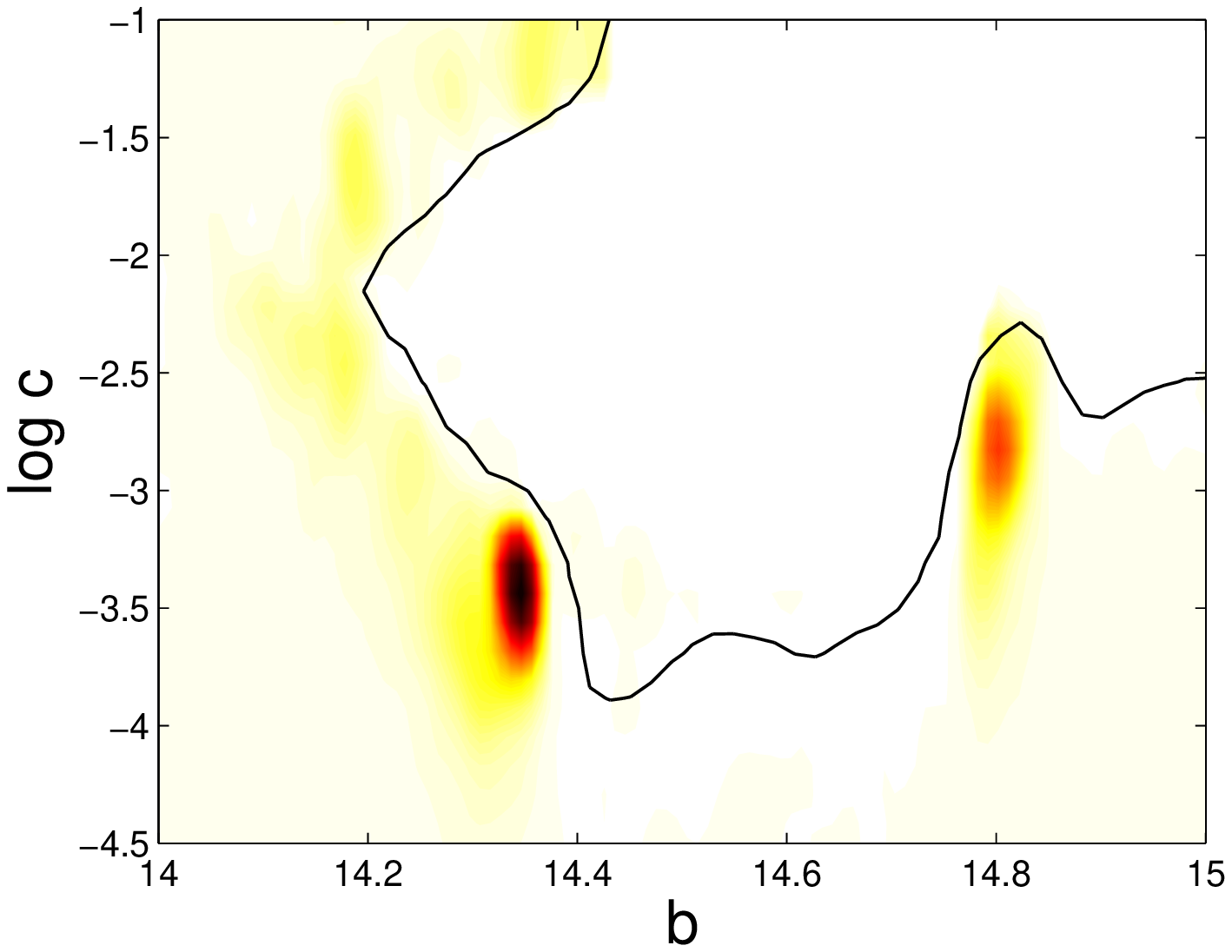}
	\includegraphics[width=.4\textwidth]{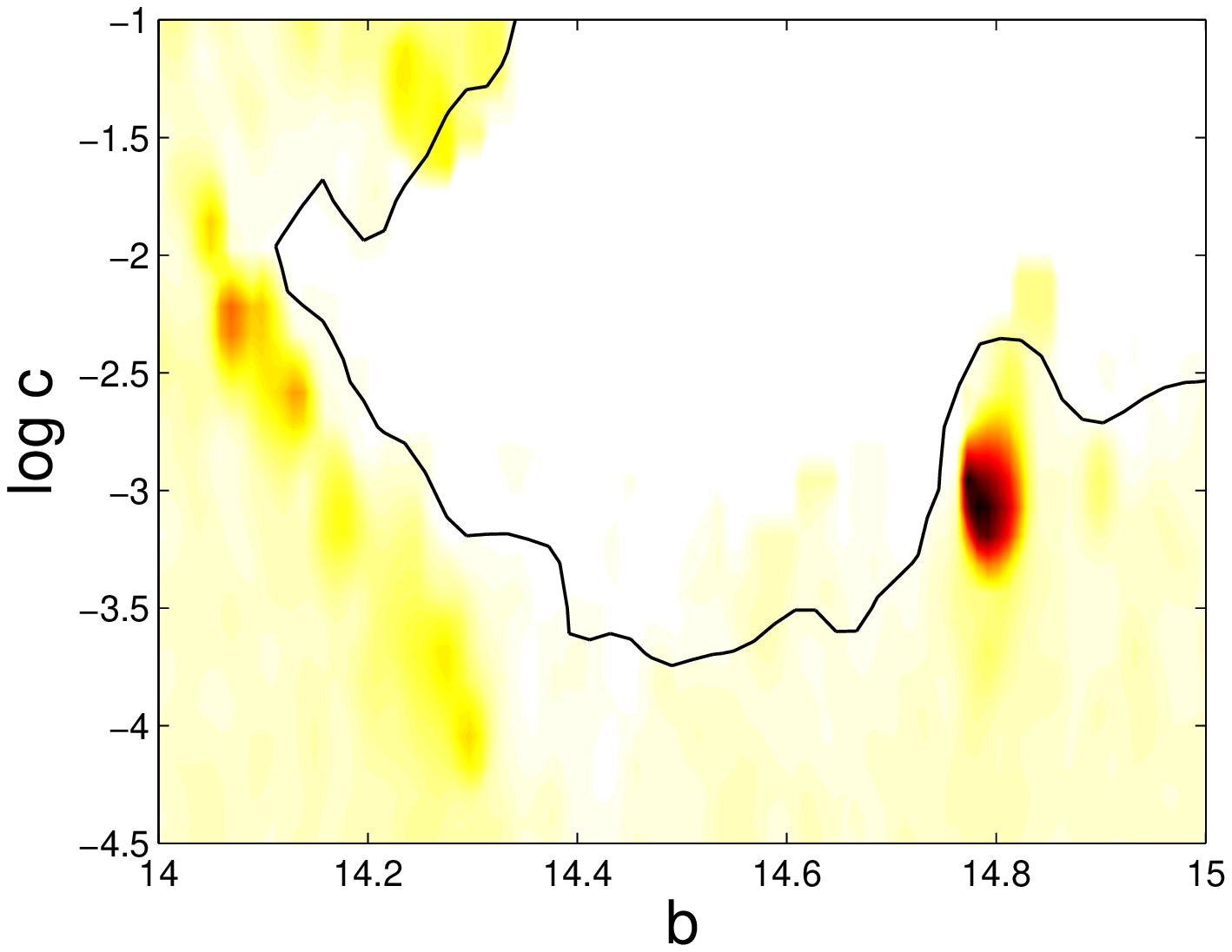}
	\includegraphics[width=.4\textwidth]{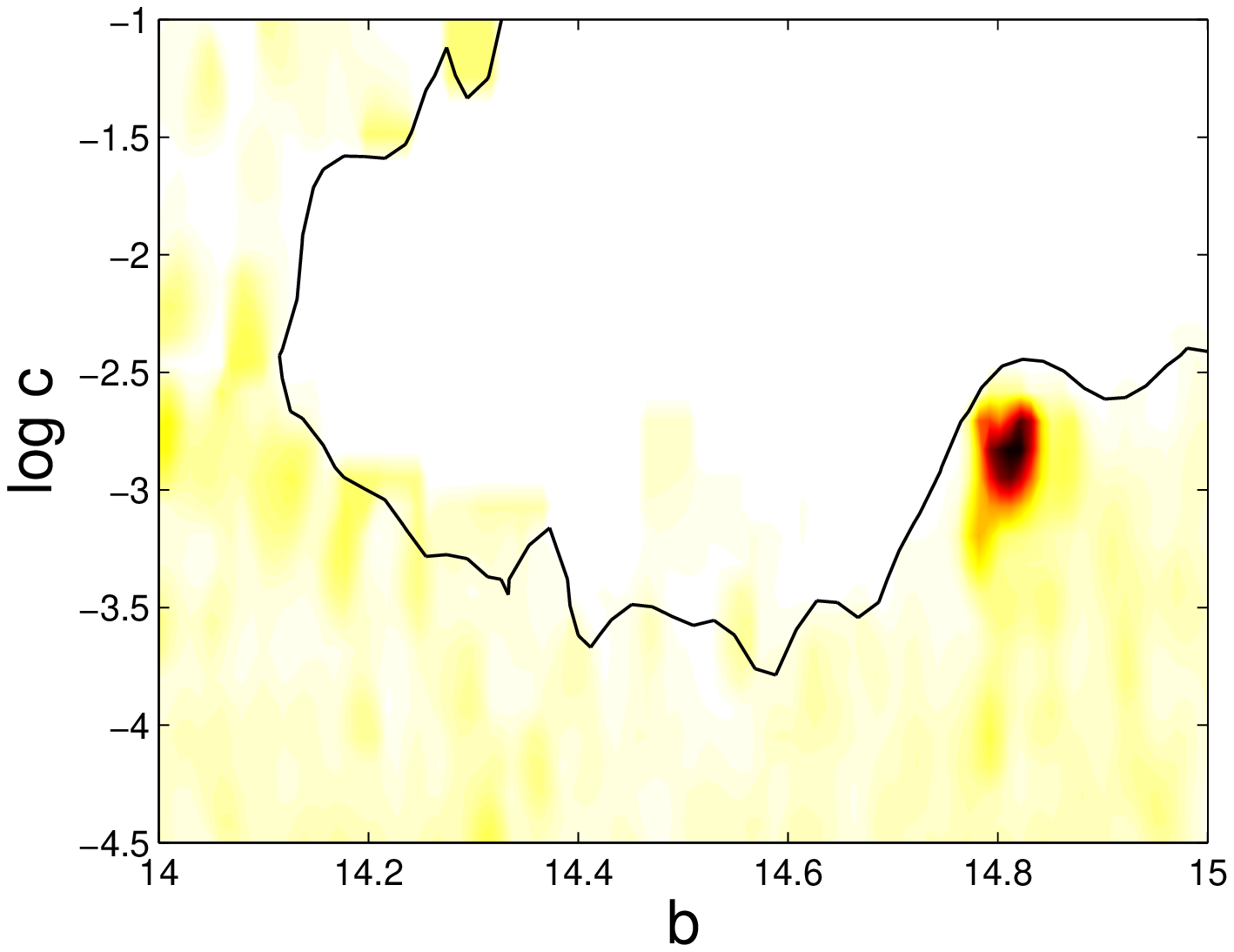}
	\caption{99\% confidence level contours for model B in the
	($b$, $\log c$) plane of parameter space with data sets 1
	(top), 2 (centre) and 3 (bottom). In these directions of
	parameter space, the likelihood function has a plateau towards
	vanishing step heights where the model reduces to the
	featureless $m^2 \phi^2$ case, an excluded valley
	corresponding to large steps and a peak at $b \simeq
	14.8$. For the WMAP data alone we also find a second peak near
	$b \simeq 14.3$.
	\label{bc}}
\end{figure}

In the present work, we replaced the old main sample data with the
luminous red galaxy sample of the most recent SDSS data release.
With this newer data set, however, we do not find such an
enhancement of the $\Delta \chi^2$ anymore. In fact, it appears to
disfavour a large feature near the third peak. Given the better
quality of the LRG power spectrum data and the fact that the BAO data
also does not seem to support this effect, it is likely that the
improvement in the fit was just a fluke. The disappearance of this
maximum of the likelihood function is illustrated in Fig.~\ref{bc},
where we show the mean likelihood (colour-coded) and the 99\%
confidence level of the marginalised likelihood in the ($b, \log c$)
plane of parameter space. The inclusion of large scale structure data
and BAO data considerably tightens the constraints on features at
small scales corresponding to values of $b$ between $\sim\!14.1$ and
$\sim\!14.4$, while for larger values of $b$, i.e., features at larger
scales, the contours remain roughly the same.

The feature at large scales ($b \simeq 14.8$), on the other hand, remains
untouched when we add the small scale data sets. For the
generalised step model and data set 1, the large scale feature maximum
likelihood point is at (\mbox{$b=14.8$}, \mbox{$c=0.001$},
\mbox{$d=0.02$}, \mbox{$\Omega_bh^2 = 0.0216$}, \mbox{$\Omega_c h^2 =
0.102$}, \mbox{$\tau = 0.11$}, \mbox{$n_\text{S} = 0.952$},
\mbox{$\log [ 10^{10} A_\text{S} ] = 3.05$} and \mbox{$H_0 = 72.7$}),
which lies near the maximum of the marginalised 1D likelihoods of the
vanilla model in Fig.~\ref{1dmarge}. This is a further indication
that the presence of a feature at large scales will not affect the
estimates of the other parameters.

Going from model B to the generalised step model will slightly improve
the quality of the fits, yielding an extra $\Delta \chi^2$ of
1-2. We did not expect a major improvement here, since the spectral
tilt of the $m^2 \phi^2$ model lies fairly close to the best fit value
of the vanilla model with a freely varying $n_\text{S}$.

\begin{figure}[!t]
	\subfigure{\includegraphics[width=.23\textwidth]{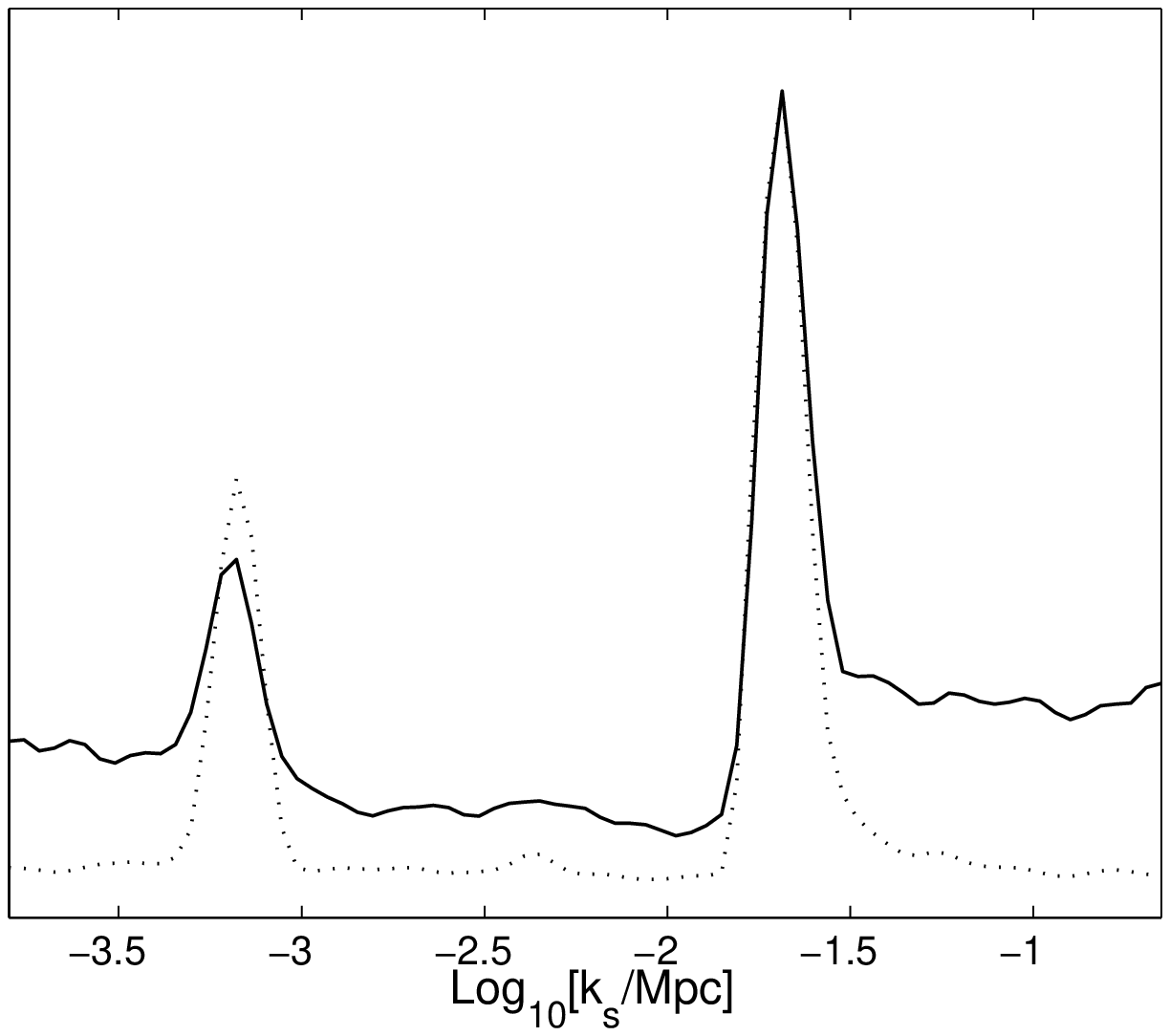}}
	\subfigure{\includegraphics[width=.23\textwidth]{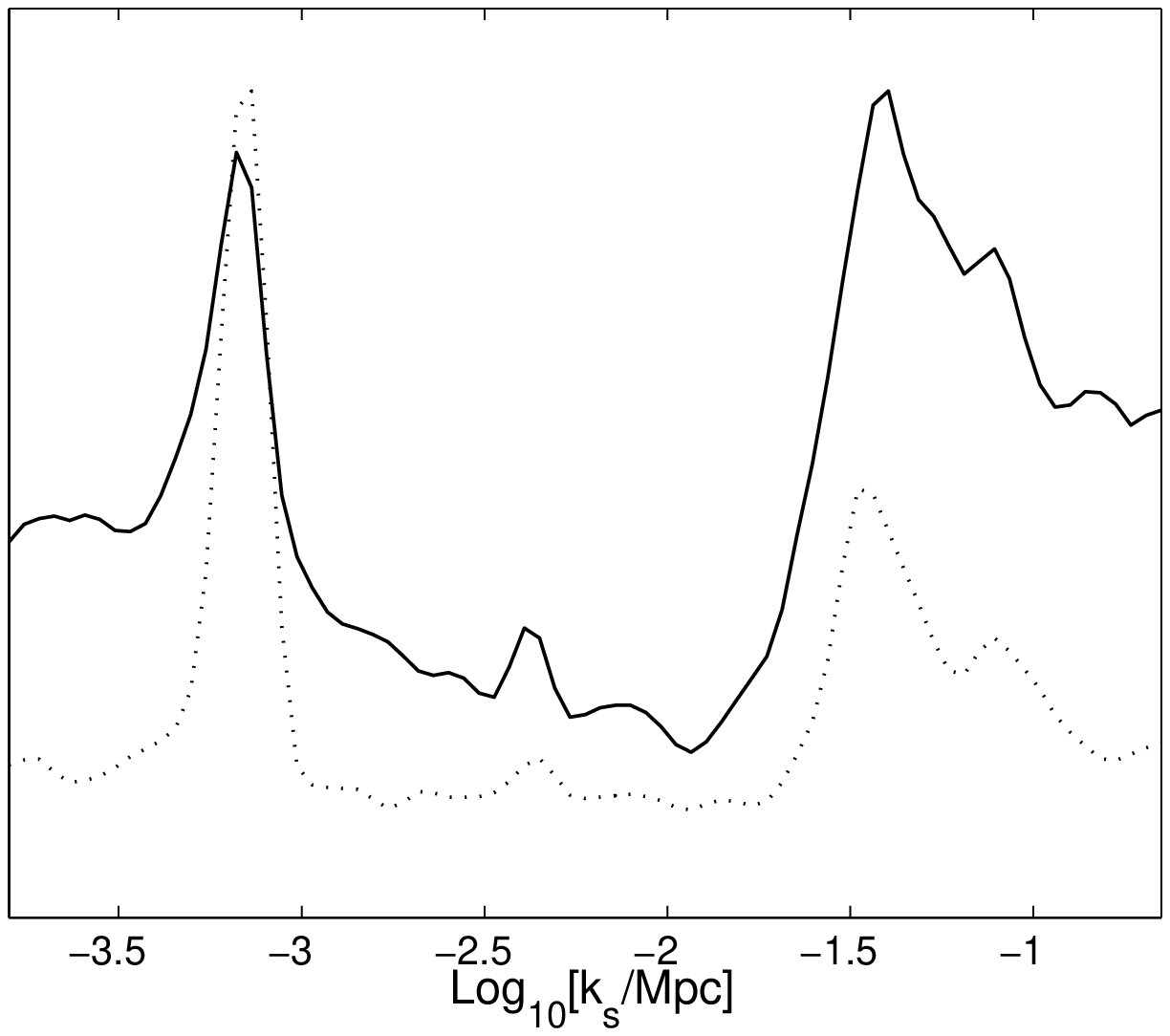}}
	\subfigure{\includegraphics[width=.23\textwidth]{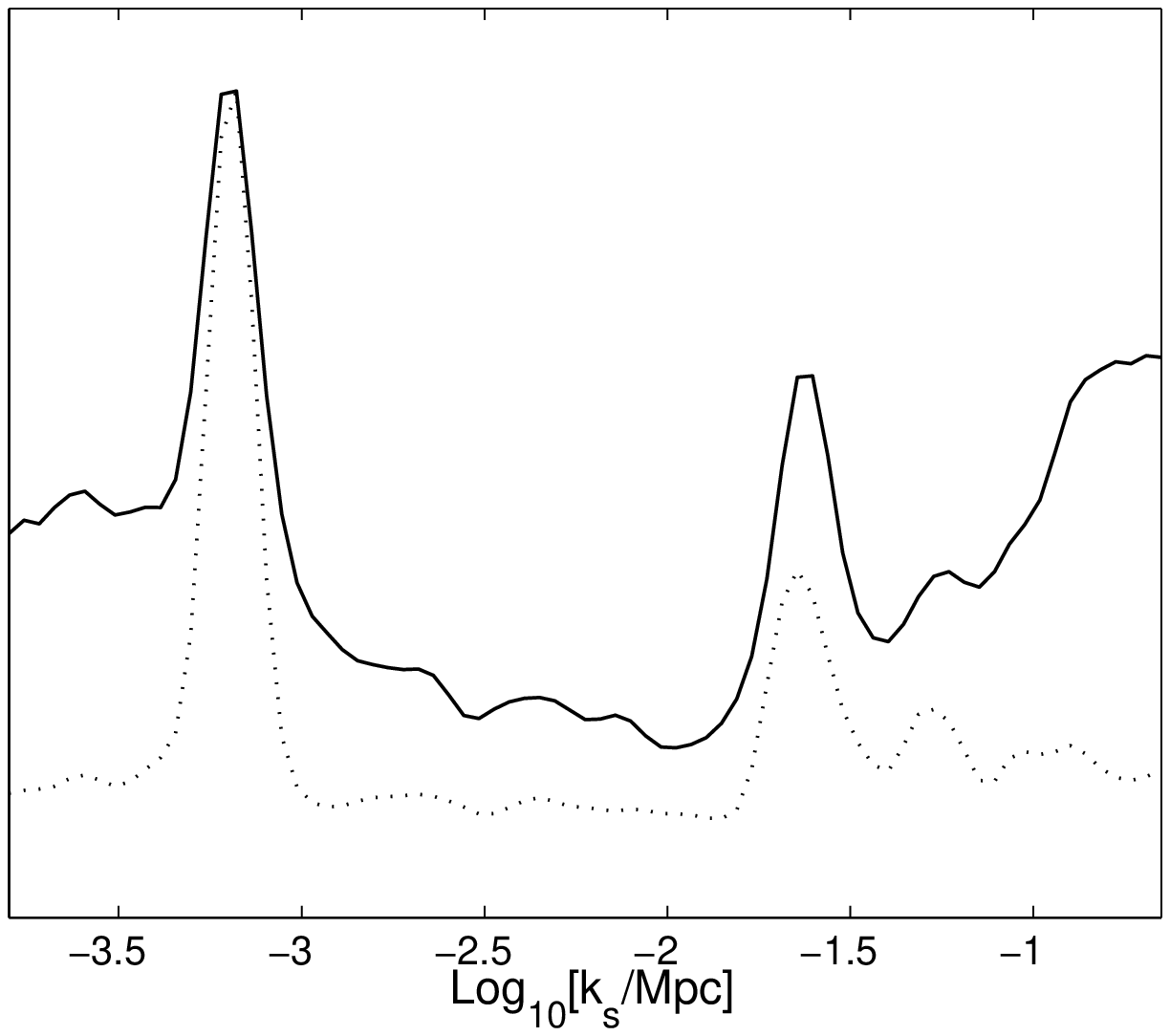}}
	\caption{Marginalised (solid line) and mean likelihoods
	(dotted line) for parameter $k_\text{s}$ in the generalised
	step model, indicating at which wavelengths a feature is
	likely to happen. Top left: data set 1, top right: data set 2,
	bottom: data set 3.
	\label{b1d}}
\end{figure}

In Fig.~\ref{b1d}, we show the marginalised and mean likelihoods for
the wavenumber $k_\text{s}$ of the perturbations that left the horizon
when the inflaton field passed the step (i.e., at the moment when
$\phi = b$). Again, we can see how the inclusion of the data sets
sensitive to smaller scales reduces the evidence for a feature at
scales \mbox{$\gtrsim \mathcal{O}(10^{-2}) \; \text{Mpc}^{-1}$}. The
difference between the mean and marginalised likelihoods is due to a
volume effect: integration over the low $c$ plateau of the likelihood
function tends to suppress peaks in the marginalised likelihood, which
show up more clearly in the mean likelihood.

\begin{figure}[!h]
\includegraphics[width=.5\textwidth]{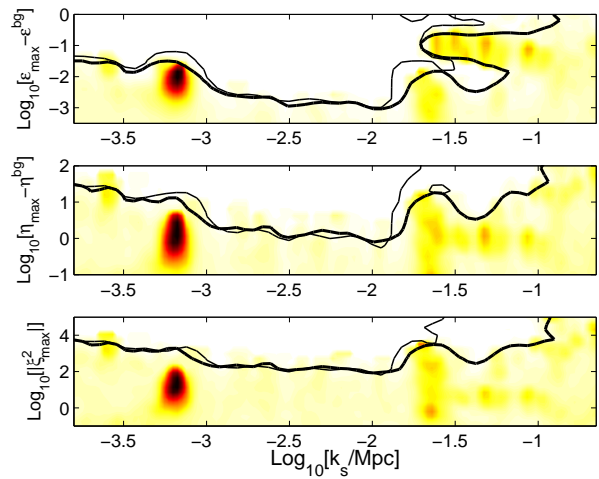}
	\caption{This plot shows the constraints on the peak values of
	the slow roll parameters during the step for the generalised
	step model. The thick line denotes the 99\% confidence level
	for data set 3, the thin line corresponds to data set 1.
	\label{slorofig}}
\end{figure}

Finally, we display constraints on the maximum values of the slow roll
parameters of the step function in Fig.~\ref{slorofig}. While the
WMAP3 data alone is only sensitive to features up to a wavelength of
\mbox{$\sim 10^{-2}\; \text{Mpc}^{-1}$}, the large scale structure data
extends the sensitivity by almost a factor ten in $k$. We find fairly
strong bounds on the maximum value of $\epsilon$ for the step
function. In conjunction with Eq.~\eqref{tensorstuff}, this implies
that the spectrum of tensor perturbations is unlikely to experience an
oscillatory modulation like the scalar spectrum, since that would
require $\epsilon$ to be of order one.

For the higher order slow roll parameters, values up to a few (for
$\eta$) and up to a few hundred (for $\xi^2$) are still allowed. Note,
however, that these bounds are parameterisation dependent (they assume
a $\tanh$-form of the step), and, for $\eta \gtrsim 1$, not only
$\xi^2$, but also higher order potential slow roll parameters will be
non-negligible.

%%%%%%%%%%%%%%%%%%%%%%%%%%%%%%%%%%%%%%%%%%%%%%%%%%%%%%%%%%%%%%%%%%%%%%%%%
%%%%%%%%%%%%%%%%%%%%%%%%%%%%%%%%%%%%%%%%%%%%%%%%%%%%%%%%%%%%%%%%%%%%%%%%%
\section{Conclusions}
\label{conc}
%%%%%%%%%%%%%%%%%%%%%%%%%%%%%%%%%%%%%%%%%%%%%%%%%%%%%%%%%%%%%%%%%%%%%%%%%
%%%%%%%%%%%%%%%%%%%%%%%%%%%%%%%%%%%%%%%%%%%%%%%%%%%%%%%%%%%%%%%%%%%%%%%%%
We have analysed the dynamics of single-field inflation models with a
step-like feature of small amplitude in the inflaton
potential. Generically, the resulting spectrum of scalar perturbations
will resemble that of the stepless background model with a
superimposed burst of oscillations whose shape is determined by the
form of the step only. We have confronted the theoretical predictions
for the spectrum of a specific chaotic inflation model with a step
with recent cosmological data to find out whether the data require the
presence of such a feature and whether it may actually bias the
estimates of other cosmological parameters such as, e.g., the baryon
density. We have also repeated the same analysis for a more empirical but
less model dependent spectrum, such as might be expected from a step in
an arbitrary inflationary background model.

With a combination of different data sets, a large chunk of the step
model parameter space can be ruled out, only spectra with a very modest
oscillation amplitude are still consistent with observations. The BAO
data, in particular, prove to be a very sensitive probe for oscillating 
spectra. 

Compared to the 6 parameter ``vanilla'' cosmological model, using the
most constraining data set, we find an improvement of the
best fit $\chi^2$ of about $5$ for the chaotic inflation step model
which comprises two extra parameters, and $\Delta \chi^2 \simeq 7$ for
the generalised step model, which has three extra parameters.

The vanilla model is a subset of the class of generalised step models
for $c \rightarrow 0$. If $c \lesssim \mathcal{O}(10^{-5})$, the
resulting spectrum will be virtually indestiguishable from the vanilla
spectrum. With our choice of priors, contours of greater than $\sim 20
\%$ confidence level will contain parts of this vanilla region of
parameter space. Hence, we cannot exclude the vanilla model at more
than 20\% confidence level. Reversing the argument, the present data
do not show compelling evidence for requiring a spectrum with an
oscillatory feature of the type discussed above.
We expect that a more sophisticated model selection analysis along the
lines of Refs.~\cite{Parkinson:2006ku,Liddle:2006tc,Liddle:2007fy}
would lead to a similar conclusion.

The best fit region of parameter space consists of models which
show oscillations at wavelengths corresponding to multipoles $\ell
\simeq \mathcal{O} (10)$, where the temperature-temperature
correlation data of the CMB shows some glitches. Interestingly, the
time it would take the inflaton field to traverse the step in these
models is of the order of an $e$-folding, which is what one would
expect for the time of a phase transition in more realistic
multi-field models.

Whether the glitches are just statistical flukes or stem from a
physical effect, such as a feature in the inflaton potential, cannot
be conclusively decided until we have better measurements of the $E$-
and $B$-mode polarisation spectra from experiments like PLANCK
\cite{planck} or, in the more distant future, projects like the
Inflation Probe \cite{inflationprobe}. An additional consistency check
can be provided by an analysis of the bispectrum of CMB fluctuations,
since the interruption of slow-roll may also induce sizable
non-Gaussianities \cite{Chen:2006xj}.

%%%%%%%%%%%%%%%%%%%%%%%%%%%%%%%%%%%%%%%%%%%%%%%%%%%%%%%%%%%%%%%%%%%%%%%%%
%%%%%%%%%%%%%%%%%%%%%%%%%%%%%%%%%%%%%%%%%%%%%%%%%%%%%%%%%%%%%%%%%%%%%%%%%
\acknowledgments
%%%%%%%%%%%%%%%%%%%%%%%%%%%%%%%%%%%%%%%%%%%%%%%%%%%%%%%%%%%%%%%%%%%%%%%%%
%%%%%%%%%%%%%%%%%%%%%%%%%%%%%%%%%%%%%%%%%%%%%%%%%%%%%%%%%%%%%%%%%%%%%%%%%
We thank Yvonne Wong for comments and discussions.
We wish to thank Irene Sorbera for valuable discussions during the
initial stages of the project. LC and JH acknowledge the support of
the ``Impuls- und Vernetzungsfonds'' of the Helmholtz Association,
contract number VH-NG-006.

%%%%%%%%%%%%%%%%%%%%%%%%%%%%%%%%%%%%%%%%%%%%%%%%%%%%%%%%%%%%%%%%%%%%%%%%%
%%%%%%%%%%%%%%%%%%%%%%%%%%%%%%%%%%%%%%%%%%%%%%%%%%%%%%%%%%%%%%%%%%%%%%%%%

%%%%%%%%%%%%%%%%%%%%%%%%%%%%%%%%%%%%%%%%%%%%%%%%%%%%%%%%%%%%%%%%%%%%%%%%%%%%%%%%%%%%%%%%%%%%%%%%%%%%%%%%%%%%%%%%%%%%%%%%%%%%%%%%%%%%%%%%%%%%%%%%%%

\end{document}